# Towards control of steady state plasma on Tore Supra

P. Moreau, O. Barana, S. Brémond, J. Bucalossi, E. Chatelier, E. Joffrin, D. Mazon, F. Saint-Laurent, E. Witrant and Tore Supra Team

*Abstract*—**Magnetic Fusion Research worldwide is now, with ITER, about to demonstrate the scientific feasibility of fusion energy production. Feedback control of fusion experiment gets more and more crucial both for performance, stability and machine protection. The Tore Supra tokamak is well suited to tackle these issues due to its unique capability to perform long duration discharges with many actuators/sensors available. The Tore Supra real time measurements and control system has been upgraded to address schemes dedicated to long pulse operation with simultaneous control of an increasing number of plasma parameters. A review of recent progress on several key control issues like measurement drift during long pulses, high efficient fuelling, plasma current profile tailoring, plasma facing component protection and self plasma protection is given.**

## I. INTRODUCTION

Achieving long-duration high performance feedback controlled discharges in a magnetic fusion device is one of the most important challenges to prepare the operation of fusion reactor [1], [2]. Hence, most of the major new projects on fusion energy, planned or under construction (W7-X, HT7-U JT60-SC, KSTAR, SST-1, and ITER) share this aim. Tore Supra (TS) tokamak is the largest superconducting magnetic fusion facility (torus dimensions: $R = 2.40\ m$, $a = 0.72\ m$, plasma current $I_p \leq 2\ MA$ and magnetic field $B_t \leq 4.0\ T$). It has been devoted to long-duration high-performance discharge research. Recently, TS went through a major upgrade replacing all the in-vessel components by actively cooled components aiming at increasing its pulse duration ability. In 2002, discharges up to 6 minutes 24 seconds duration with injected / extracted energy up to 1 GJ have been performed. That offers a unique capability of addressing the plasma control issues in long pulse operation towards steady state plasma control.

The plasma may be modelled as a resistive ionised fluid moving in a magnetic field. It reacts as a multi-time scale, non-linear distributed system with a large number of potential instabilities. Plasma parameters are often strongly coupled and available actuators are still in limited number. They consist in an external set of magnetic coils, pellet and gas injection, and heating systems. Plasma control has to be performed at several physics time-scale connected to different physical processes (Fig. 1): typically 10-100ms for plasma equilibrium, and plasma fuelling, a few seconds for plasma current diffusion, tens of seconds to minutes for plasma wall interaction. For intrinsically unstable and complex system such as confined plasma, feedback control clearly has a crucial role for performance optimisation and machine protection.

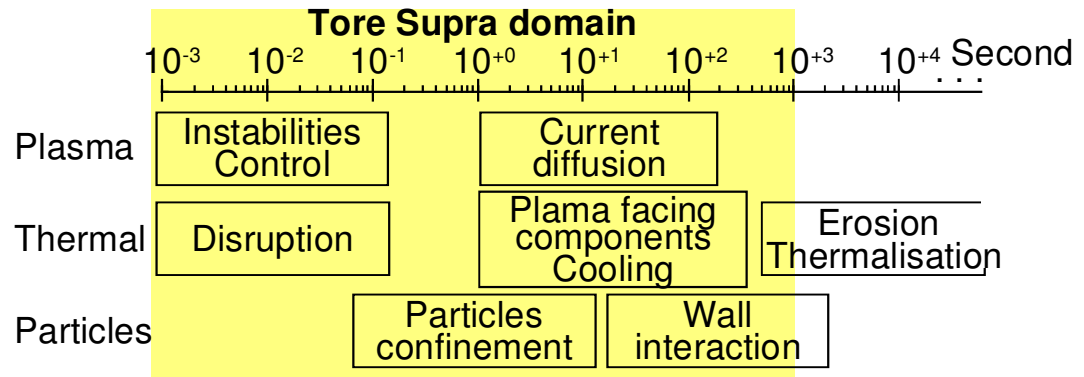

Fig. 1: Characteristic time scale on fusion devices.

This paper is an overview of basics and recent progress on TS real time measurements and control system. Section II describes the hardware. It depicts the network systems used for diagnostics (sensors), real time (RT) data computation, feedback controller and actuators. Section III discusses main key control issues: plasma equilibrium, plasma fuelling, plasma internal profiles, plasma facing components protection and pulse management. Section IV gives a conclusion pointing out the future needs.

## II. HARDWARE

Most TS diagnostics use an acquisition unit equipped with two processors, each in charge of a specific function. The first one is dedicated to the communication with the real time server in order to synchronize the acquisition and the control with the timing unit of the discharge, transmit raw and processed data and store them. The second processor runs a single RT task dedicated to acquisition on input boards, raw data processing, using control loops of a few milliseconds from a specific algorithm. It is used to deliver the calculated control voltages to actuators or subsystems it manages. Intercommunication between processors is achieved by a Versatile Module Eurocard (VME) bus through shared memory.

Recently, PC units (INTEL Pentium® IV-2.8GHz) have been used for RT computation: high level feedback controller and plasma equilibrium reconstruction are now routinely available.

The TS data acquisition system must fulfill a broad variety of requirements. First, continuous data acquisition has been implemented, meaning that the supervision storage and timing tasks are continuously running at low sampling frequency. This allows continuous data recording of some diagnostics like calorimetric sensors, which is of major importance for plasma facing components heat load studies.

In the opposite, some data acquisition units require a high data flow rate (100 kHz up to 1 GHz) when special plasma event occurs. During 1-2 seconds, several times per discharge, the data flow rate can reach 18 Mb/s per front-end unit. For these units, the row data are transferred via a private 100 Mb/s Ethernet link to separate powerful PC units, where the data are computed and sent to the central RT server to be stored in the database. Using such a technique, a pseudo-real time calculation can be implemented into the PC to achieve feedback control at a somewhat lower frequency.

Finally, a multi-parameter integrated RT control of the plasma requires information coming from many diagnostics. The one to one connection between sensors and actuators of the initial control topology is no longer sufficient to fulfill these requirements. Sharing information, of measured quantities and computed parameters as well becomes an essential issue. Therefore a fast dedicated network has been built (SCRAMNet® boards from SYSTRAN Corporation), connecting control units together (Fig. 2).

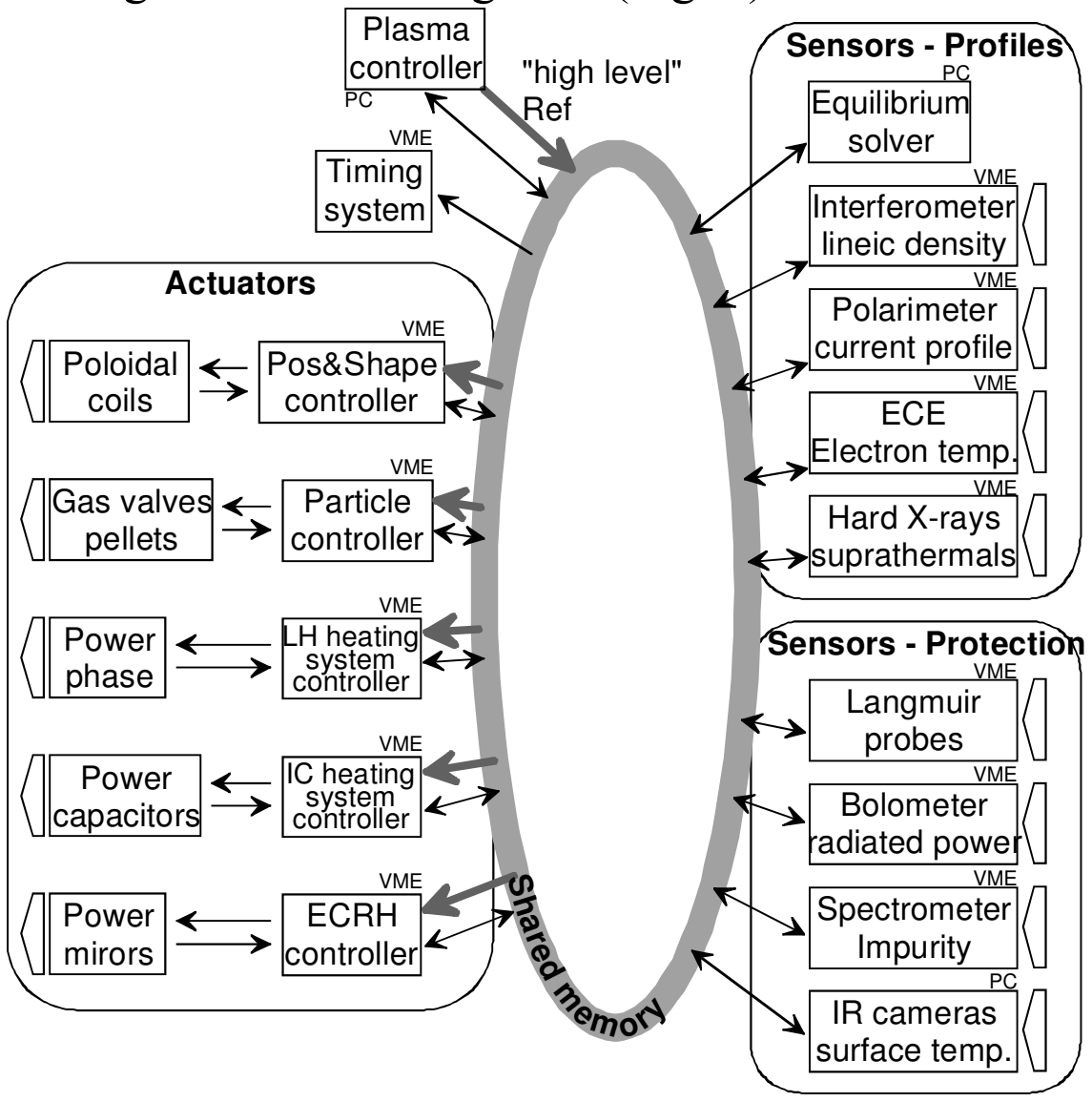

Fig. 2: Data acquisition system of Tore Supra and RT capabilities

RT sharing of information ensures a global and consistent sub-units operation. This shared memory is now routinely used on TS to perform an accurate plasma control. A central control unit collects the information from all diagnostics and calculates "high level" references which are sent to the actuators through the shared memory.

## III. Recent progress on steady state plasma control

### A. *Plasma equilibrium*

TS poloidal field system [3] fulfills in a single set of coils the ohmic heating and the plasma position and shape control. It consists of nine coils connected to nine independent power supplies used to control the plasma current and the magnetic configuration. The generator $G_0$ controlling the central solenoid (A coil in Fig. 4) can be used either to drive the plasma ohmic current or to fix the plasma flux at the last closed flux surface for zero loop voltage operation. The height remaining generators are used to control the plasma shape and position.

The TS plasma position and shape controller uses 51 measurements $B_\theta^m$ of the poloidal magnetic field (pick-up coils), 51 measurements $B_\rho^m$ of the radial magnetic field (pick-up coils) and one toroidal flux loop measuring the poloidal flux. The pick-up coils are located on a circular shaped surface taken as the reference surface (Fig. 3). During inductive phases, the current is induced into the plasma by transformer effect using the central solenoid (A coil) as primary. The structure of the toroidal pumped limiter is a conducting ring. Therefore, current is also induced in this structure in the ratio of the plasma resistance and the ring support resistance. A model of the influence on the pick-up coil of this current flowing into the toroidal pump limiter structure has been developed and the measurements are corrected in RT from this current.

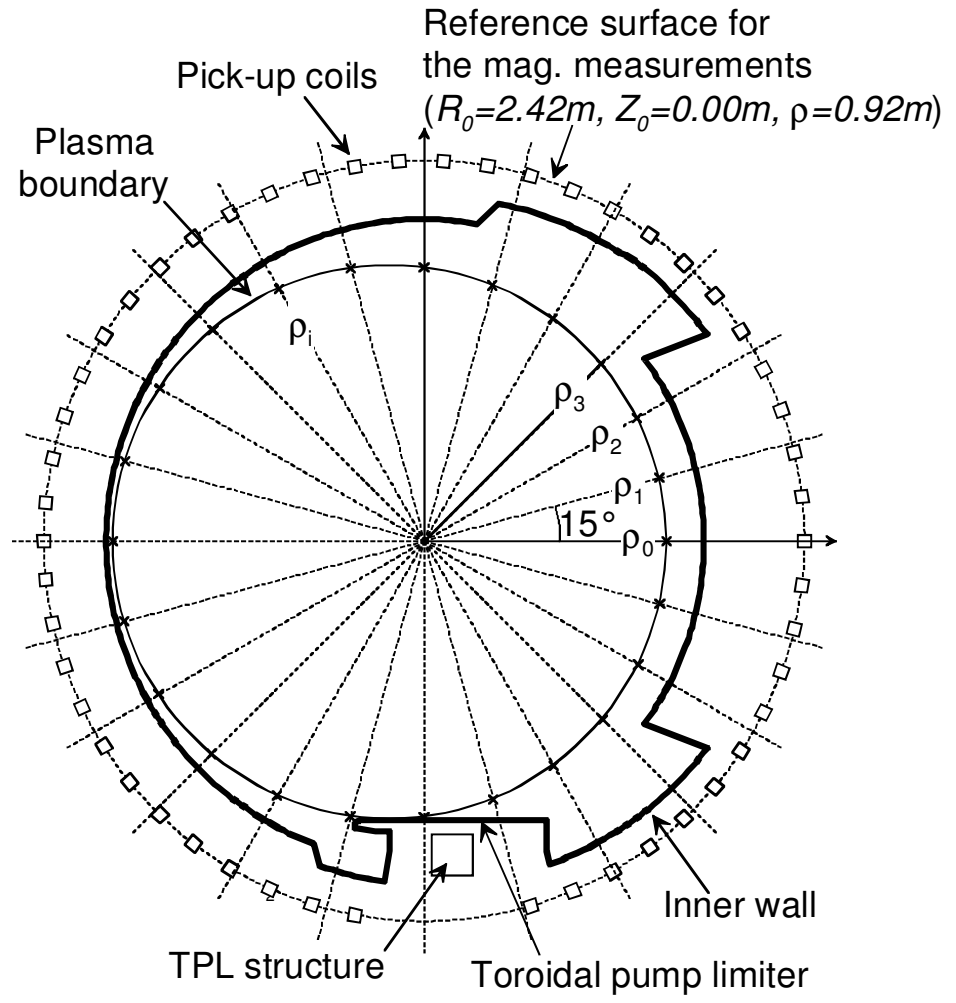

Fig. 3: Poloidal section of Tore Supra showing the magnetic probe positions (square), the reference surface (dotted line circle), the inner first wall including movable limiters (full line), Toroidal pump limiter structure and the 24 control axis (dashed line).

The plasma boundary is given by the last closed isoflux surface. Using the magnetic measurements and a Taylor expansion, the plasma flux is calculated at the intersection between 24 predefined radial directions (control axis $\Delta\theta=15°$ centred on the reference surface) (Fig. 3) and the first wall. The first wall is defined by the position of the movable limiters and the inner first wall after a geometric correction for toroidal ripple effect. The largest flux point is then considered as the contact point of the plasma to the first wall and the corresponding flux is the plasma flux $\psi_{plasma}$. 24 radial distances $\rho_j$ along the control axis are derived from the isoflux contour $\psi_{plasma}$. Finally, 24 radial distances difference between the desired and the actual plasma boundaries, measured along the control axis, is obtained by:

$\Delta\rho_j = \dfrac{\psi_j - \psi_{plasma}}{(\partial\psi/\partial\rho)_j}$ where $\psi_j$ and $\partial\psi/\partial\rho$ are calculated at 24 predefined control axis. A feedback control matrix **F** converts the 24 $\partial\rho_j = \Delta\rho_j - \overline{\Delta\rho_j}$ into eight voltages to be delivered by the poloidal field generators. A proportional integral (PI) controller is used with global weighting factors *G* and *I* (Fig. 4):

$$V = \mathbf{F} \otimes \left[ G\,\partial\rho + I \int_0^t \partial\rho\, dt \right]$$

$\partial\rho$ quantities are used rather than $\Delta\rho$ to be insensitive to the contact point. **F** is an 8 by 24 matrix where coefficients have been defined theoretically [4] and adjusted empirically using open loop experiments. Absolute errors for the plasma major radius $R_p$ and vertical position $Z_p$ are within 2 mm fulfilling the requirements. The control loop cycle is 2 ms, with typical CPU time (VME 300MHz-PowerPC unit) 1.8 ms, including data reading and saving (0.4 ms) calibration (0.2 ms), boundary solver (0.9 ms), feedback (0.2 ms) and safety control (0.1 ms). This is consistent with the 8ms PF system response time.

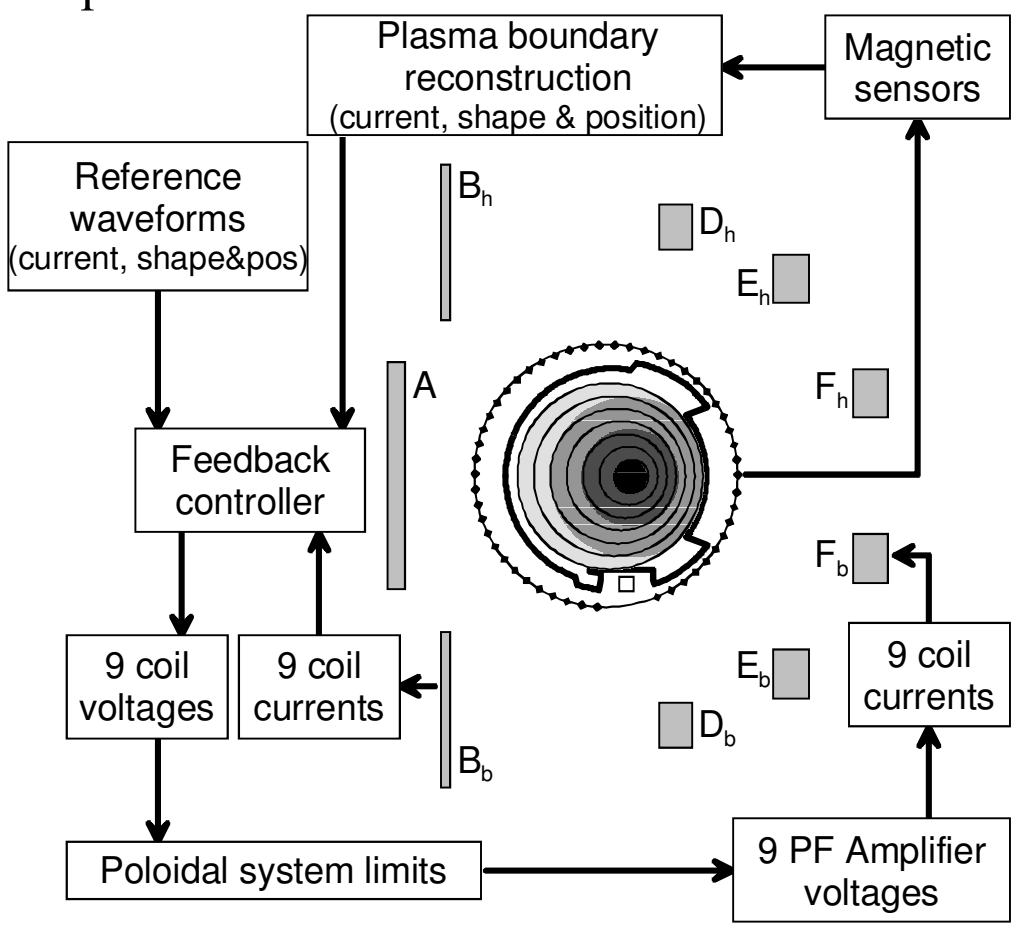

Fig. 4: Plasma current, position and shape feedback loop.

The major issue towards steady state is related to the magnetic sensor accuracy. During long pulse operation, the electronic of magnetic sensors, which mainly consist of integrators, is subject to drifts, affecting the position and shape control. Even if the integrator drift can be reduced [5], equilibrium reconstruction techniques that could cope with it are highly desirable. Such technique has recently been developed on TS. It consists in modulating the plasma position in both directions ($R_p$, $Z_p$), and the plasma current $I_p$. This adds new information to those available from probes in a static equilibrium. Explicit modulation of the plasma current is necessary, since the plasma radius modulation influences the current (and vice versa) and it is vital to separate these two effects. By demodulating the magnetic sensors data, it is shown that amplitude and phase behavior strongly depending on plasma position (Fig. 5). Another way consists in identifying the contributions of $R_p$, $Z_p$, $I_p$ in the sensors data modulation [6]. In both cases, the plasma position can then be identified by neural network techniques.

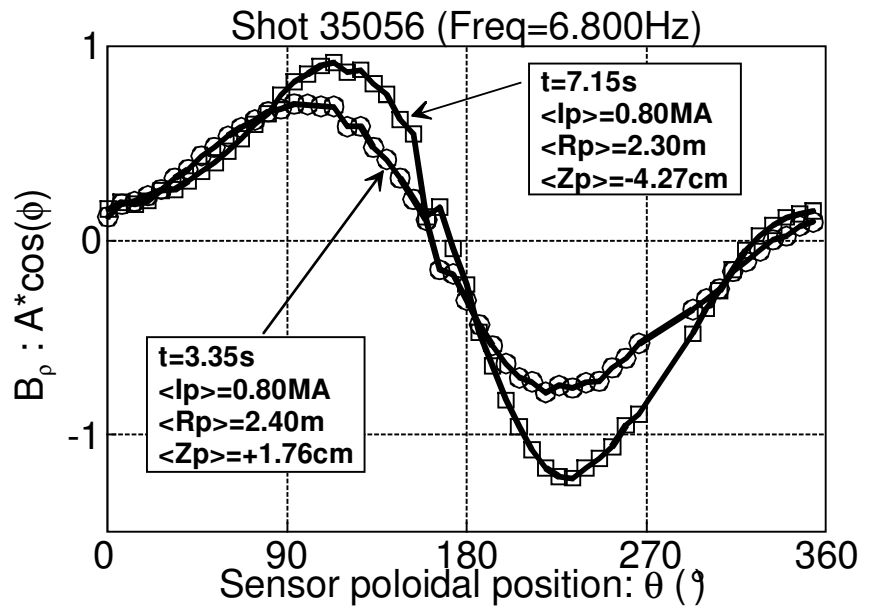

Fig. 5: Magnetic sensors amplitude and phase calculated for two different plasma positions (square and circle).

## B. *Plasma fuelling*

The particles control is an essential issue in long duration plasma discharges. TS is provided by a unique set of fuelling namely: gas puffing, supersonic molecular beam injection and pellet injection. All of them have RT capabilities for density feedback control during long pulse operation.

The gas puffing is the basic tool to control the plasma density in tokamaks. It requires very little hardware (piezoelectric valves) and is very reliable. But the fuelling efficiency (10-20%) is low compared to the other techniques because the gas is ionised at the plasma edge. In TS, a PI controller ensures the gas puffing feedback. Using the calibration (voltage/flow rate) of the piezoelectric valves, the controller calculates the voltage to be applied to the valves:

$$V = G\,\delta Nl + I \int_0^t \delta Nl\, dt$$

where $\delta Nl = Nl^{(meas)} - Nl^{(ref)}$ is the difference between the reference and the measurement. *G* and *I* are the weights of the PI controller. Several valves can be used at the same time with possibly different types of gas.

With supersonic gas injectors it is possible to launch a series of very short (2ms) and dense gas jets at Mach number 5. This system exhibits a better fuelling efficiency (40-50%) than the gas puff. Although the edge plasma is strongly perturbed during the gas pulses (nearly detached phase of ~40 ms), ion cyclotron (IC) and lower hybrid (LH) additional power can still be coupled to the plasma. The feedback controller is simple: when the measured density is lower than the reference, the gas controller asks for a gas pulse injection via the TS timing system. This operation can be repeated at several Hertz (up to 10Hz) to maintain the request of density. Fully supersonic gas injection fuelled plasmas have been successfully tested during 60s pulse discharges [7].

Pellet injection is the most promising technique in particular for next step facility like ITER due to its better efficiency (100%) which should allow minimising particle in-vessel retention. It consists in injecting pellets of deuterium into the plasma. The set-up is technically more complex and TS is provided with an injector that can inject cylindrical pellets (diameter of 1.7 or 2 mm) continuously at a frequency up to 10 Hz and a velocity between 100-600 m/s, with a very high reliability (~ 99%). The feedback controller acts in the same way as for the supersonic molecular beam injection. Simultaneous pellet fuelling and coupling to the plasma IC and LH additional power is a real challenge. On one hand, suprathermal electrons driven by LH waves prevent pellet from getting deep into the plasma and on the other hand, perturbation of edge density by pellet injection may prevent IC power to be coupled to the plasma. Thus, additionally, each pellet is preceded by a notch of IC and LH power 30 ms before it enters into the plasma. Pellet fuelled LH driven discharges lasting up to 2 minutes have been performed [8]. One hundred and fifty-five pellets have been injected into the plasma from the low field side, at a frequency close to 1.3 Hz under feedback for maintaining the line density near the target value of $2.5x10^{19}$ $m^{-3}$ (Fig. 6).

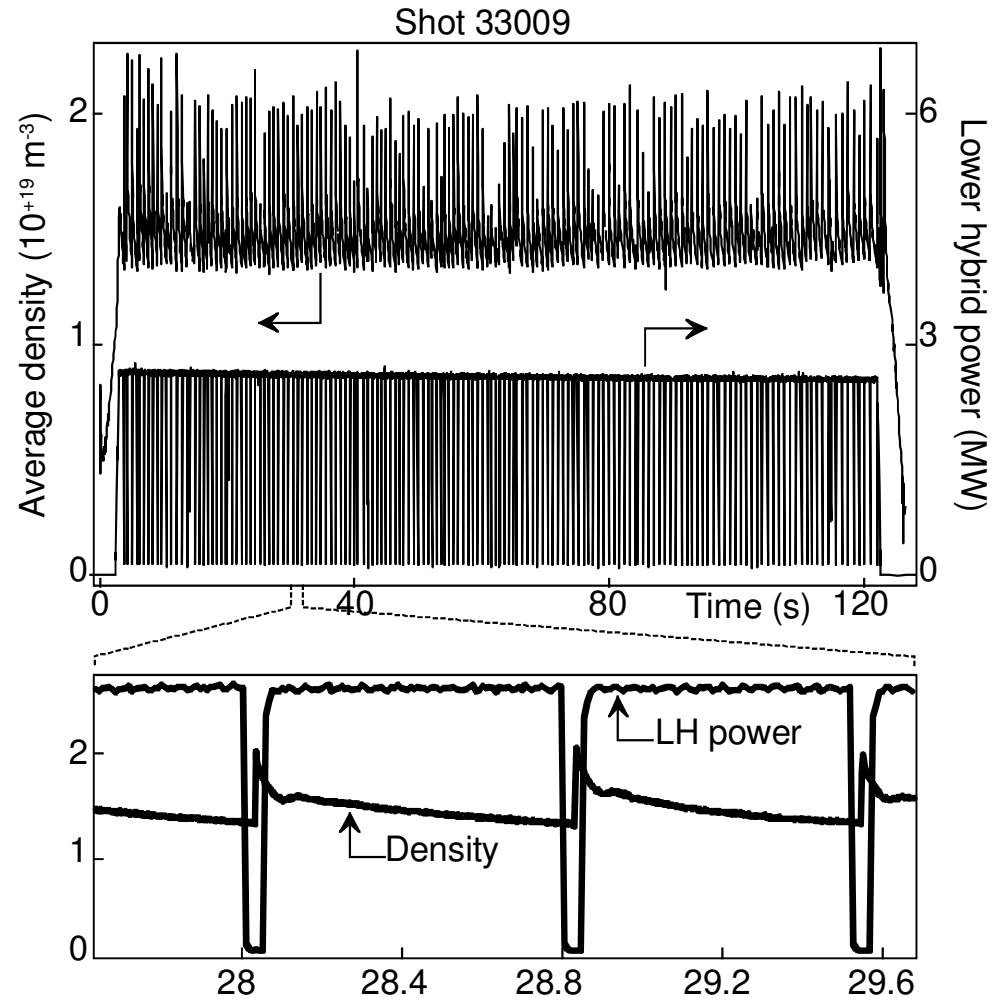

Fig. 6: Time evolution of the plasma density and lower hybrid power for a 2 minutes plasma discharge fully fuelled by the pellets.

The gas controller must be as robust as possible with respect to measurements. The RT density measurement is performed by an infrared interferometer which could be subject to fringe jump during very fast plasma change thus giving the wrong density value. The TS controller does detect such events and switches automatically to the density given by the Bremsstrahlung diagnostics with a small loss of precision (Fig. 7). The gas controller also has safety role. For example when the radiated fraction approaches 90%, the plasma detaches from the wall. This high radiation regime is usually not compatible with the RF waves coupling, and is prone to disruptions. Therefore the gas injection is stopped until the fraction of radiated power comes back below a given threshold (typ. 70%). This feedback control has proven to be extremely efficient to prevent disruptions.

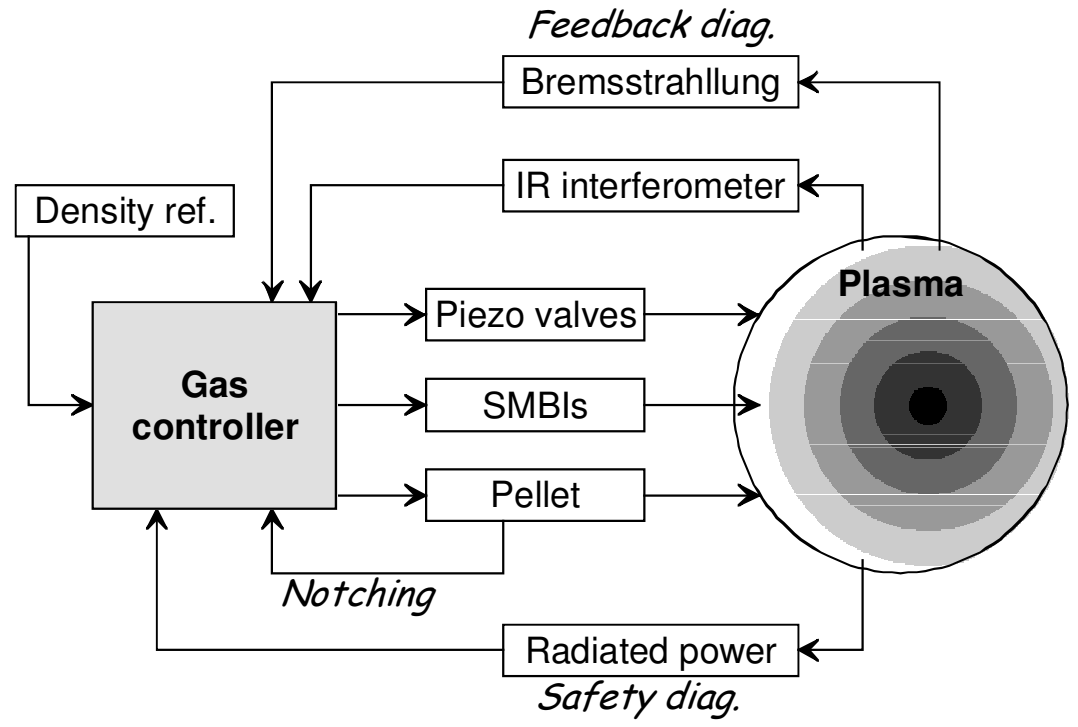

Fig. 7: Block diagram of the gas controller.

## C. *Plasma profile parameters*

The local plasma parameters (i.e. plasma parameters profile) control has become an important issue on the way to high performance, instability free steady state plasma discharges. This field requires powerful diagnostics for RT profile computation, more sophisticated controller taking into account the profile shape, actuators having the ability to modify locally plasma parameters and local plasma models.

In TS, LH additional power is the dominant external source used for non-inductive discharges. Thus the LH power deposition profile is strongly linked to the generated current profile. The measurement of the Bremsstrahlung radiation emission in the hard X-rays range by the suprathermal electrons generated by the LH waves is the most effective method to get information about the LH deposition profile. Using the RT signal of the hard X-ray diagnostics, a feedback control of the current density profile has been performed in TS [9].

As a starting point in the direction of controlling the plasma current profile, the width at half maximum of the hayd X-ray emission profile is used. Two actuators have been studied: the parallel refractive index $n_{//}$ of the injected LH wave and the LH power $P_{LH}$. The dependency of the profile's width on both actuators has been determined experimentally: increasing $n_{//}$ increases the profile's width [10] and the LH power acts in the same direction. The PI controller weights have been calculated, in a first step, from open loops Taylor discharges giving controller static gains and, in a second step, they have been adjusted from the analysis of close loop plasma discharges.

In TS present non-inductive discharges, the plasma current is fully sustained by LH waves. Achieving fully non-inductive discharges requires the control of the central solenoid flux consumption using the $G_0$ power supply, simultaneously with the non-inductive control of the plasma current using the LH power. Such feedback control is routinely operated in TS for long duration discharges [1], [2], [11], [12]. In parallel, the current profile control can now be ensured using the LH refractive index $n_{//}$ as actuator.

The plasma parameters are strongly coupled (Fig. 8) even though, in a first step, the controllers have been developed neglecting this coupling.

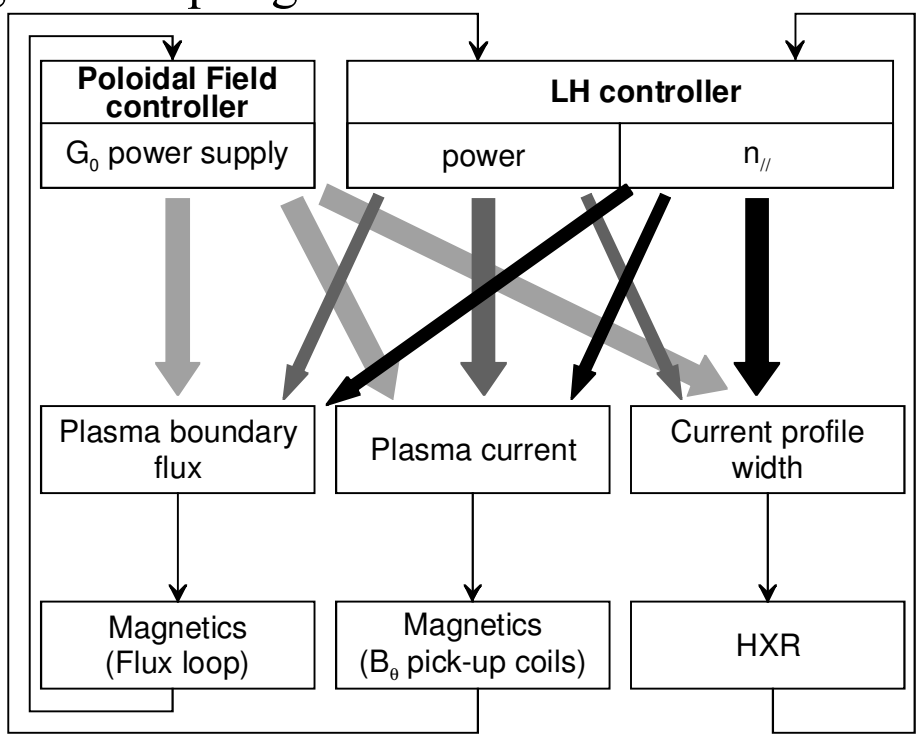

Fig. 8: Example of plasma parameters coupling and feedback controllers. The width of arrow is related to the coupling intensity.

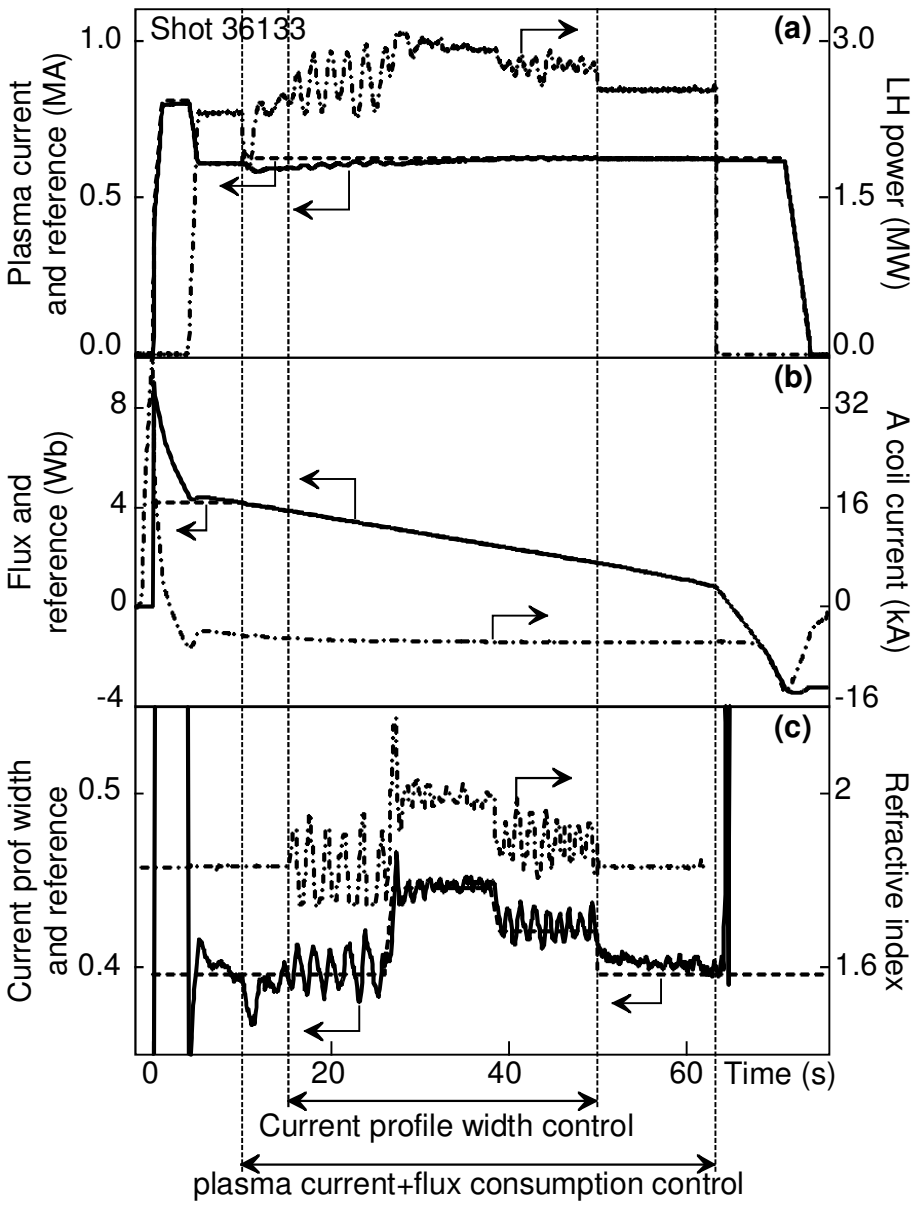

Fig. 9: Multiple control operating simultaneously: (a) plasma current control, (b) flux consumption control and (c) current profile width control. The solid line is the measurement, the dashed line the reference and the mixed line corresponds to the actuator.

It is important to note that the plasma controller does not calculate a new reference level but modulates the existing reference using a coefficient varying within 0 and 1. In that way, the LH protection is ensured because the variation of $n_{//}$ or LH power is bounded. The demonstration of a distributed coupled parameters feedback control achieving (i) plasma current control from LH wave power, (ii) flux consumption from $G_0$ power supply and (iii) the current profile width from the LH refractive index is shown Fig. 9.

### D. *Plasma facing components protection*

During operation of present fusion devices, plasma facing components (PFCs) are submitted to large heat fluxes. Understanding and preventing overheating of these components during long pulse discharges is a crucial issue for next step tokamaks, in particular to avoid damage or undesired erosion of the components.

An infrared thermography diagnostics has been implemented on TS as a part of the CIMES project [13]. The monitoring of the most sensitive components, namely 3 IC antennae, 2 LH launchers and the toroidal pumped limiter is performed in RT.

While the toroidal pumped limiter has been designed to sustain heat flux of 10 MW $m^{-2}$ at steady state, the most critical points are antennae and launchers, where hot spots or overheating of large areas can be observed during high-injected power plasma discharges. Critical areas have been identified on each antennas and launchers. The analysis of the heating processes identified the role of the private power (HF sheaths or fast electrons) and the cross interactions area between antennas and launchers (fast ions or fast electrons) (Table I).

TABLE I

DESCRIPTION OF INTERACTIONS BETWEEN IC AND LH ANTENNAE.

| Area of interest | Interaction | Mechanism | Feedback controller |
|---|---|---|---|
| LH launcher – Guard limiter inner parts | LH → LH | Fast electrons generated in front of the LH launcher | Reduce the power of the incriminated LH launcher |
| LH launcher – Wave guide below mid-plane | IC → LH | Fast ions generated by IC wave | Reduce the total IC power |
| IC antenna – Guard limiter | LH → IC | Fast electrons generated in front of the LH launcher | Reduce the total LH power |
| IC antenna – Faraday screen | IC → IC | RF sheaths in front of IC antenna | Reduce the power of the incriminated IC antenna |

Using the RT thermography diagnostics, a feedback control has been implemented to prevent components overheating. Prior to the shot, areas of interest are selected on the PFCs and a physical interaction process is associated to each of them (private power or cross interaction with other heating system). During the shot, the maximum temperature is calculated in each area of interest and sent to central plasma controller unit which, decides whether the power has to be reduced and which heating system the reduction is applied on. The feedback control is seen as a hybrid controller in the sense that it is activated only if a temperature of a selected area of interest approaches selected threshold (Fig. 10).

Such control has been successfully validated on Tore Supra. Moreover, the compatibility with other feedback controls like zero loop voltage or the width of the current profile has been demonstrated (Fig. 11). The control of the PFC temperature is ensured simultaneously with the control of the current profile width using the refractive index of the LH system. As we can see (Fig. 11), the target profile width between 24-30s and 39-46s are the same. This target has been reached by the controller even though the LH power has been stepped down at 32s due to over heating of the launcher. This results in slight increase of the LH refractive index.

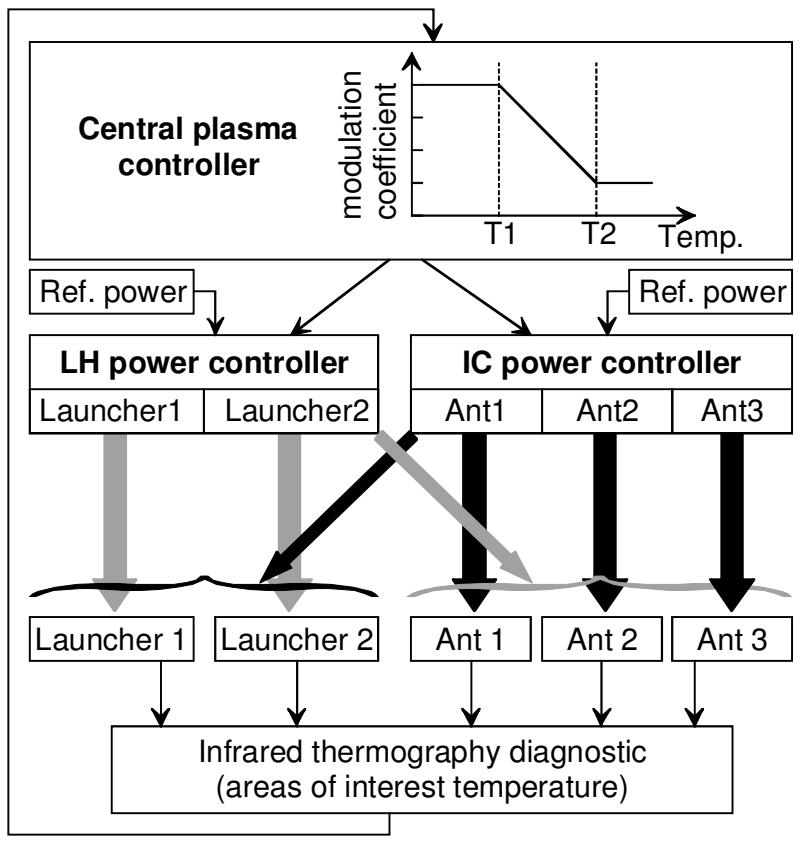

Fig. 10: principle of power reduction to limit plasma facing components overheating.

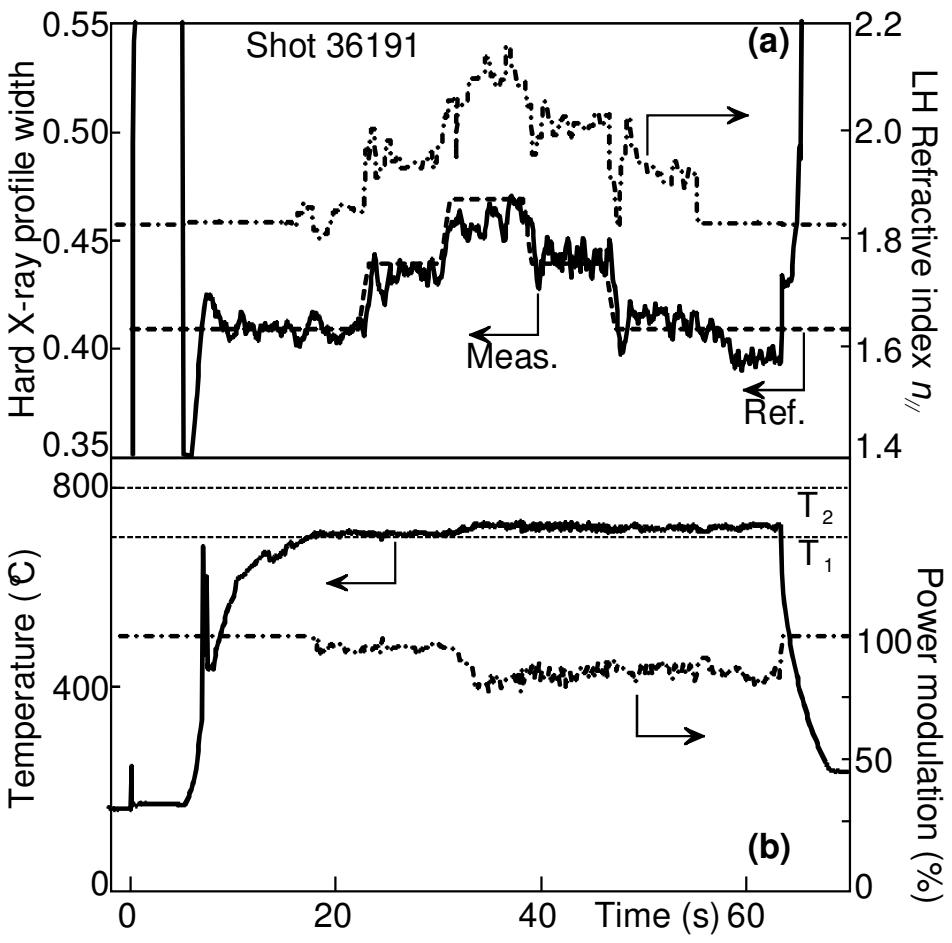

Fig. 11: Example of compatibility between feedback controls: (a) control of hard X-ray profile width using LH refractive index, (b) Temperature limitation decreasing the LH power.

The feedback control can be used to optimize the additional heating operation while keeping the plasma facing components temperature within their operational limits.

### E. *Plasma pulse termination control*

Disruptions are a major problem for tokamaks operation. During such event, forces up to hundred tons can be applied to structures and a significant fraction of the plasma current can be converted into fast electrons (50MeV). Massive gas injection technique is used on TS to reduce disruption impact. Encouraging tests in have been carried out recently [14]. Disruption predictor has been derived with a good level of confidence. It combines RT magnetic instabilities data (pick-up coils) and fraction of radiated power computed by bolometer diagnostics. When these two quantities increase over an experimentally adjusted threshold, massive gas injection is triggered. In parallel, heating power and plasma fuelling are stopped and weights of the PI plasma equilibrium controller are slightly decreased in order to keep the control of the plasma position until no current is detected.

## IV. Conclusion

Feedback control is a central tool to optimize the plasma performance and safety. While global parameters are successfully controlled for basic operation, the steady state high performance operation brought to light new challenges in plasma profile control, plasma stability management and power exhaust control. The short term challenge is to integrate all these controls in a single controller. Already, long duration discharges characterized by simultaneous current profile, plasma equilibrium, flux consumption and plasma facing components temperature controls have been performed in TS. These new fields of investigation require model based controllers, taking into account the multi-time scale distributed non linear nature of the underlying physics. Very challenging work is still required in this matter for present and future tokamaks like ITER.